# Graphene Based Waveguide Polarizers: In-Depth Physical Analysis and Relevant Parameters

Rafael E. P. de Oliveira* and Christiano J. S. de Matos

MackGraphe – Graphene and Nanomaterials Research Center, Mackenzie Presbyterian University, São Paulo, 01302-907, Brazil
*rafael.oliveira1@mackenzie.br

**ABSTRACT**

Optical polarizing devices exploiting graphene embedded in waveguides have been demonstrated in the literature recently and both the TE- and TM-pass behaviors were reported. The determination of the passing polarization is usually attributed to graphene's Fermi level (and, therefore, doping level), with, however, no direct confirmation of this assumption provided. Here we show, through numerical simulation, that rather than graphene's Fermi level, the passing polarization is determined by waveguide parameters, such as the superstrate refractive index and the waveguide's height. The results provide a consistent explanation for experimental results reported in the literature. In addition, we show that with an accurate graphene modeling, a waveguide cannot be switched between TE pass and TM pass via Fermi level tuning. Therefore, the usually overlooked contribution of the waveguide design is shown to be essential for the development of optimized TE- or TM-pass polarizers, which we show to be due to the control it provides on the fraction of the electric field that is tangential to graphene.

## Introduction

Graphene is a one-atom thick carbon allotrope arranged in a crystalline hexagonal structure that presents remarkable properties such as a high electronic mobility and linear electronic dispersion in which the valence and conduction bands touch at a single momentum-space point[1]. The latter characteristic leads to a constant 2.3% absorption when light, over a broad optical band, traverses a single graphene layer at normal incidence[1,2]. Also, nonlinear optical effects in graphene are notably high[3], with a reported nonlinear susceptibility[4] of $\chi^{(3)}=1.4\times10^{-15}$ m$^2$V$^{-2}$ and a broadband saturable absorption[5]. These unique properties make graphene an attractive constituent for the development of high-speed and broadband photonic devices, such as modulators and polarizers[1,6].

Even with the optical absorption per unit length being remarkably high in graphene, the net absorption it provides is low due to its atomic thickness. For this reason, optical devices that incorporate graphene usually enhance the light-graphene interaction by embedding it longitudinally, close to the core of waveguides[2]. Using this geometry optical modulators[7-10], and polarizers based on both optical fibers[11,12] and integrated waveguides[13,14] have been demonstrated.

The polarizing effect in graphene-waveguide structures occurs because optical absorption in graphene takes place only for the fraction of the electric field that is parallel to graphene's plane[7], and because this fraction is generally different for transverse-electric (TE) and transverse-magnetic (TM) radiation (with TE and TM being defined relative to the plane of incidence for reflection in the surface where graphene is placed). Bao *et al.* demonstrated a TE-pass polarizer in the optical fiber geometry, yielding up to 27 dB extinction ratio at 1550 nm wavelength, that used a ~3-mm length of graphene[11]. A similar geometry was exploited by Lee *et al.* using a 5-mm graphene length[15], with extinction ratios of 0.07 dB, with air above graphene, and 3.36 dB, with a liquid over graphene, being reported. However, in contrast to the results reported by Bao *et al.*, a TM-pass characteristic was observed.

Polarizing effects were also reported in integrated waveguides. Kou *et al.* measured the polarization dependent loss in integrated silicon waveguides with graphene and reported a TM-pass behavior with an extinction ratio of 40 dB/mm around 1550 nm[13]. Kim and Choi demonstrated a polymeric integrated waveguide which changes its polarizing characteristic from TE (~10 dB extinction ratio) to TM pass (~19 dB extinction ratio) when a dielectric upper cladding is added, covering the waveguide with graphene[14]. The graphene length was ~7 mm and the measurements were performed at 1310 nm wavelength.

The operation of each specific reported device as TE pass or TM pass is usually explained in terms of graphene's Fermi level. In some cases, the excitation of waves similar to surface plasmon polaritons (SPP) is said to occur at doping levels for which Pauli blocking takes place and intraband transitions dominate over the interband transitions[11-14]. In such cases graphene's conductivity is predominantly imaginary and positive, and the TM radiation is said to be guided in graphene with low loss, resulting in a TM-pass polarizer. In the absence of Pauli blocking, the devices are said to behave as TE-pass polarizers[11-14].

This description of the guided light-graphene interaction has led to proposals of switchable polarizers[2,12] controlled by a gate voltage[16], but such devices have not yet been demonstrated experimentally[2]. Indeed, the TM-pass devices reported in the literature[13,14] present no specific method for doping graphene and it is unlikely that accidental doping would provide the required Fermi level displacements (e.g., a >0.4 eV displacement would be required for a TM-pass polarizer at 1550 nm). In addition, no specific method for coupling light to the mentioned SPP-like mode is reported, which is usually required for plasmon-polariton excitation due to phase mismatch[17]. Furthermore, a rigorous model describing the proposed SPP-like propagation in the reported polarizers is not provided.

Here we show through numerical simulations that the waveguide design, rather than graphene's doping level or coupling to SPP-like waves, determines the passing polarization in graphene-waveguide optical polarizers and that the refractive index of the superstrate (i.e. the layer beyond graphene) and the waveguide dimensions are crucial for the development of optimized optical polarizers. In addition, we show that the TE-pass to TM-pass transition cannot occur with the tuning of graphene's Fermi level, but that it indeed takes place via tuning of the superstrate's index or of the waveguide height.

## Modeling, Results and Analysis

As several graphene-waveguide polarizers with differing structures and materials have been reported, it is instructive to, first, analyze a specific polarizer, in which the influence of different parameters can be analyzed. Therefore, simulations were initially performed by modeling a silicon nitride waveguide, a classic silicon photonics material that is of great interest for the development of graphene-based optical and electro-optical devices[18]. The simulations were then extended to different waveguides reported in the literature, with consistent results obtained. The simulations were performed in the COMSOL Multiphysics software, by two-dimensional modal analysis using the finite element method. The waveguide was designed for operation at ~1550 nm wavelength (with single mode operation being achieved with an upper cladding refractive index in excess of 1.45), which is used in the majority of the reported experiments and is of major interest for telecommunication systems. Figure 1 shows the waveguide, which has a core refractive index, $n_1$, of 2 (silicon nitride), a width (along the $x$ axis), $W$, of 1.6 μm, and a height (along the $y$ axis), $H$, of 0.8 μm. The waveguide is surrounded by a medium with refractive index, $n_2$, equal to 1 and lies on the top of a substrate with refractive index, $n_{sub}$, of 1.45 (silica). A single layer of graphene is placed on the top of the waveguide core and covers its whole width. Above graphene, a superstrate with a refractive index $n_{top}$ and a 300-nm thickness is modeled. Figure 1 also shows the normalized time-averaged power flow distribution, $S_z$, along the propagation direction ($z$ axis), in color scale, and the electric field distribution, black arrows, for the TM-mode at a 1550-nm wavelength and for $n_{top}$=1.45. The simulation computes the complex propagation constant of the guided modes; the absorption coefficient, $α$, was then calculated from the imaginary part of this constant.

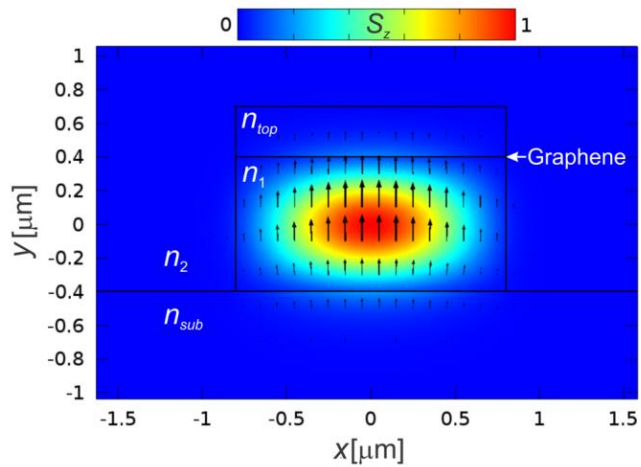

**Figure 1.** Simulated waveguide design showing the normalized optical power flow along the $z$ direction, $S_z$ (color scale), and the electric field (arrows) for the TM-mode at 1550 nm. $n_{top}$ = 1.45.

**Polarization Dependence on Graphene's Fermi level**

Graphene was initially simulated, as commonly found in the literature, as a 3D isotropic material with a thickness $t_g$=0.34 nm,[19] and a complex refractive index $n_g = \varepsilon_g^2$, where $\varepsilon_g$ is the graphene effective permittivity calculated from its sheet conductivity, $\sigma$: $\varepsilon_g = \varepsilon_\infty + i\sigma/(\omega\varepsilon_0 t_g)$, where $\omega$ is light's angular frequency; $\varepsilon_0$ is the vacuum permittivity; and $\varepsilon_\infty$ is the graphene background permittivity assumed here to be 1.[12] A background permittivity of 2.5 is used, instead, in some references,[20,21] and is found to depend on graphene's substrate[21]. Nevertheless, it is important to notice that the exact value of this parameter is not critical for the following analysis because it is real, therefore not leading to optical absorption, and does not depend on graphene's Fermi level[21].

Graphene's conductivity was calculated using the Kubo formalism, which takes into account graphene's linear electronic dispersion, including both intra- and interband transitions,[22,23] as well as temperature, graphene's electronic relaxation time, the optical angular frequency and graphene's Fermi level; these parameters were respectively set to 300 K, 500 fs,[11] ~1.215×10[15] rad/s (corresponding to the 1550-nm wavelength), and $E_F$, with the latter being varied. Figure 2(a) shows the real and imaginary parts of the calculated conductivity as functions of the Fermi level, and Figure 2(b) shows the corresponding calculated graphene permittivity.

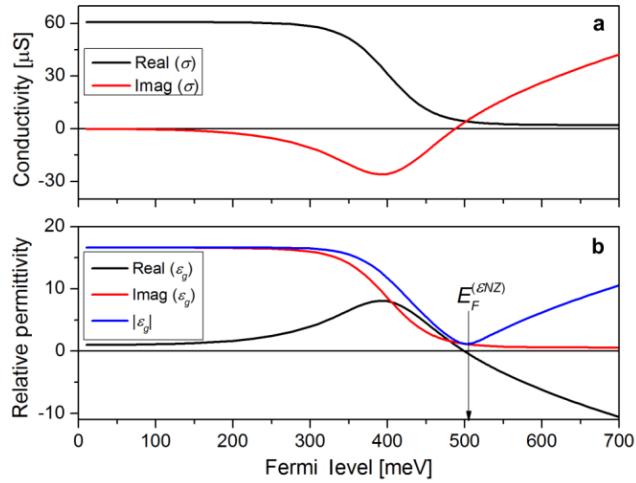

**Figure 2.** Calculated graphene's (a) conductivity, $\sigma$, and (b) effective relative permittivity, $\varepsilon_g$, as functions of the Fermi level, $E_F$, for a 1550-nm vacuum wavelength.

The photon energy at 1550 nm is $\hbar\omega$=800 meV and, as expected, in Figure 2(a) a decrease in the real part of the conductivity is noticeable for $E_F > \hbar\omega/2$ due to Pauli blocking. Below this Fermi level value, interband transitions dominate. For $E_F$ > 505 meV the conductivity becomes predominantly imaginary and the intraband transition contribution increases. The ~505 meV Fermi energy corresponds to a minimum in the absolute value of graphene's permittivity ($\varepsilon_g$ =-0.26+1.12i), which is known as the *epsilon near zero point*, *εNZ*,[24,25] the Fermi level of which is indicated in Figure 2(b). At this point the model that assumes graphene to be isotropic predicts, for the TM mode, an enhancement in the fraction of light that travels along graphene. For its field concentration within graphene, this situation may look like the excitation of surface plasmon polariton-like waves.[10] However, we instead relate such a field distribution to the electric boundary condition $\hat{\mathbf{n}} \cdot (\mathbf{D}_a - \mathbf{D}_b) = 0$ in the core-graphene interface, with $\hat{\mathbf{n}}$ representing the unitary vector normal to graphene (*xz* plane) and $\mathbf{D}_j = \varepsilon_j \mathbf{E}_j$, with *j* = *a* or *b*, being the electric displacement vector in medium *j*, near the boundary; **E** represents the electric field vector. This condition shows that the discontinuity in the permittivity, which is greater near $E_F^{(\varepsilon NZ)}$, must be accompanied by a proportional discontinuity in the normal component of the electric field. A similar discontinuity between graphene and the superstrate further enhances the normal field within graphene. The observed field enhancement is, then, similar to that reported and exploited in slot waveguides.[26] Note also that if this enhancement was related to SPP waves, there would be an enhancement in the electric field parallel to the graphene's plane, rather than an enhancement only orthogonal to the xz plane.

Furthermore, still within the isotropic graphene model, the local field enhancement would reportedly lead to an increase in the absorption for the TM mode, resulting in a TE-pass behavior around the $\varepsilon NZ$ point that is shown in Figure 3 (red curve) for $n_{top}$=1.75 and 1550-nm wavelength. Figure 3 shows the extinction ratio, defined as the difference in the absorption coefficients for the TE and TM modes $(\alpha^{(TE)}-\alpha^{(TM)})$, as a function of the Fermi level. This extinction ratio definition implies that a TM-pass (TE-pass) behavior corresponds to positive (negative) extinction ratios. The negative extinction ratio peak near the $\varepsilon NZ$ point would make possible the fabrication of polarization switchable devices controlled by a gate voltage. Polarizers[6,12] and modulators[9,20] have been proposed exploiting the $\varepsilon NZ$ point in graphene but the lack of experimental results raises questions about their feasibility.[20] Kwon shows that in the proposed modulators the $\varepsilon NZ$ effect occurs solely if graphene is considered as a 3D isotropic material,[20] but due to the confinement of the electrons in two dimensions it must be treated as an anisotropic material with the out-of-plane conductivity much smaller than the in-plane conductivity. Indeed, in this case, due to the constant out-of-plane refractive index, the normal electric field is not enhanced as explained above and the enhancement in absorption is not observed, resulting in the non-existence of the $\varepsilon NZ$ peak in Figure 3.

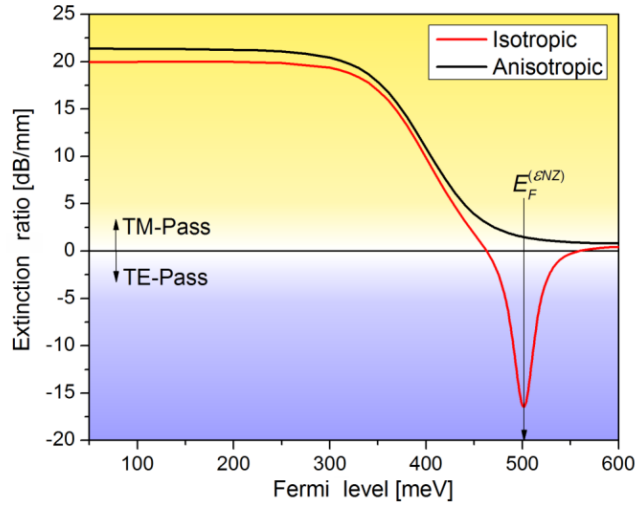

**Figure 3**. Polarization extinction ratio in the modeled graphene-waveguide polarizer, with graphene modeled as an isotropic medium and as an anisotropic medium.

Hereforth, we, thus, proceed to model graphene as an anisotropic material whose permittivity in the *y*, out of graphene's plane, direction is taken to be real and equal to 2.5, regardless of the Fermi level.[20] Alternatively, we point that it is more convenient to simulate graphene as purely bidimensional, through the inclusion of the boundary condition $\hat{\mathbf{n}} \times (\mathbf{H}_a - \mathbf{H}_b) = \sigma \mathbf{E}$ in the core-superstrate interface, where **H** is the magnetic field vector. This modeling strategy is computationally less demanding and yields negligible differences relative to the anisotropic, 0.34-nm thick, graphene model. Figure 3 also shows the extinction ratio versus Fermi level for the anisotropic graphene model (black curve), for which the TM pass to TE pass transition never occurs. Instead, a smooth decrease in the (positive) extinction ratio is observed when $E_F > \hbar\omega/2$. Fermi-level induced polarization switching was not observed in any case when $n_{top}$ was swept from 1 up to 1.8 and the waveguide height varied from 0.4 up to 1.2 μm; the same behavior was observed in simulations with the different waveguide geometries described in *Methods*. These results, therefore, rule out the proposal of polarization switching via the tuning of graphene's Fermi level in the assessed waveguide geometries.

**Polarization Dependence on the Waveguide Design**

Although the TM-pass or TE-pass behavior of the polarizer is not dependent on the Fermi level, it is determined by the exact waveguide design. Figure 4 shows the extinction ratio as a function of the superstrate refractive index, $n_{top}$, for different waveguide heights, *H*. The Fermi level was fixed at 100 meV and the quantitative results are not significantly affected by its exact value up to ~300 meV, as can be noticed in Figure 3. According to Figure 4, a TE-pass behavior is observed for low superstrate indices, with optimum extinction ratios obtained in waveguides without a superstrate ($n_{top}$=1). The TM-pass behavior is, on the other hand, achieved with higher superstrate refractive indices. However, when the waveguide height is lower than a given threshold, the TM-pass behavior is observed regardless of $n_{top}$ due to the strong confinement of light in the *y* direction, which increases the TE-mode interaction with graphene. The inset in Figure 4 shows the extinction ratio as function of the waveguide height for $n_{top}$=1; an optimized TE-pass polarizer is obtained for *H*~0.55 µm. The change in the waveguide width was also evaluated but a much smaller influence on the polarization dependent loss was noticed.

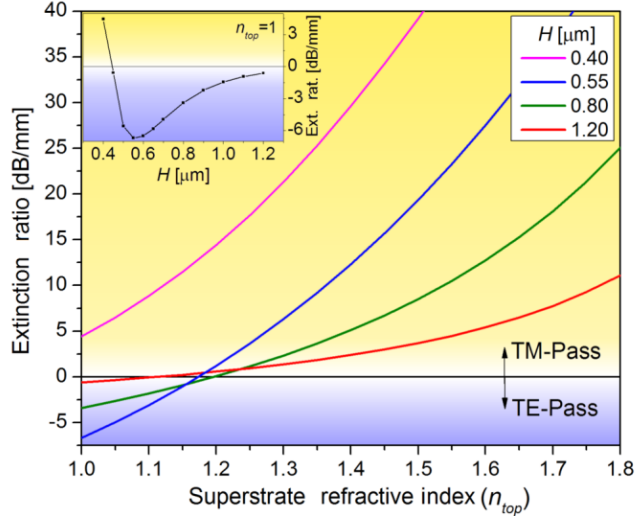

**Figure 4.** Polarization extinction ratio as function of the superstrate refractive index for different waveguide heights, *H*. *Inset*: extinction ratio as function of the waveguide height for $n_{top}$=1.

These results can be explained through the evaluation of the tangential electric field, $E_{||}$, at the graphene layer for each mode and specific waveguide design. Only this field component has an effective interaction with graphene. Therefore, the mode that presents the highest fraction of its field parallel to graphene, at the graphene position, experiences the highest absorption. For comparing different modes, we define a normalized tangential field intensity, integrated along graphene's width at the waveguide cross section:

$$U_{||} = \frac{t_g \int_{-W/2}^{W/2} |E_{||}|^2 \, dx}{\iint |S_z| \, dxdy}, \qquad (1)$$

with $|E_{||}|^2 = |E_x|^2 + |E_z|^2$. For the TE mode, $E_{||}$ is dominantly composed of $E_x$, while for the TM mode $E_{||}$ is mostly composed of $E_z$. Figure 5 shows $U_{||}$ for the TE and TM modes as a function of $n_{top}$ in a waveguide with *H*=0.55 µm. The result shown is related to the blue curve of Figure 4, for which the transition point from TE to TM pass occurs at $n_{top}$=1.17. This superstrate index is marked in Figure 5 as a vertical solid line and perfectly matches the point beyond which $U_{||}$ for the TE mode supersedes $U_{||}$ for the TM mode. Therefore, it is shown that the polarization dependent loss is, indeed, determined by the fraction of the field that is parallel to graphene and that the waveguide design directly determines this fraction.

In the extreme points of each curve of Figure 5, the distribution of the parallel electric field intensity, $|E_{||}|^2$, across the waveguide is also shown. For the TE mode and for low superstrate indices, most of the parallel electric field is concentrated around the center of the waveguide; the increase in $n_{top}$ leads to a weaker mode confinement and, therefore, to a shift of the mode distribution towards the superstrate. This trend increases the interaction of light with graphene and, consequently, increases absorption. On the other hand, for the TM mode, a low $n_{top}$ means a high core-cladding index contrast, which tends to increase the fraction of the electric field that is parallel to the propagation direction and concentrated at the edges of the waveguide; as $n_{top}$ increases, the weaker mode confinement makes the field distributions tend to those of transverse electromagnetic (TEM) modes, thus, reducing the electric field along *z*. Consequently, a reduction in the absorption of the TM mode is observed. The attenuation for the TE (TM) mode is 29.79 dB/mm (36.47 dB/mm) for $n_{top}$=1 and 58.93 dB/mm (10.50 dB/mm) for $n_{top}$=1.8.

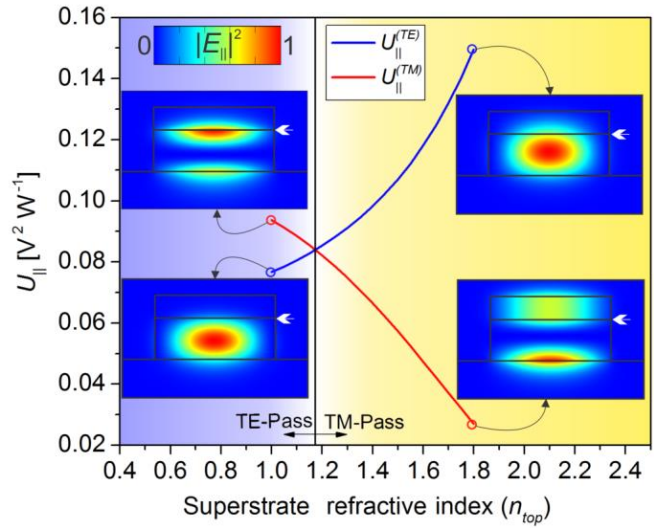

**Figure 5**. $U_{\parallel}$ as a function of the superstrate index for the TE ($U_{\parallel}^{(TE)}$) and TM ($U_{\parallel}^{(TM)}$) modes. *Insets:* parallel electric field intensity ($|E_{\parallel}|^2$) distributions across the waveguide.

## Discussion and conclusions

The results obtained here offer a consistent explanation for experimental results reported in the literature. For example, the results presented by Kou *et al.*, showing a TM-pass behavior despite the absence of a superstrate can be explained by the small height (200 nm) of the silicon waveguide used in their experiments[13]. Indeed, a simulation was performed for the exact waveguide presented by Kou *et al.* and the calculated attenuation for the TE and TM modes were 0.075 dB/μm and 0.025 dB/μm respectively, consistent with the mean experimental values reported of 0.09 dB/μm and 0.05 dB/μm.[13] This result, therefore, explains the experimental observation solely via the waveguide design, without resorting to the unconventional propagation of SPP-like modes proposed by the authors.

Kim and Choi have shown in polymeric waveguides that the addition of a dielectric superstrate changes the behavior from TE to TM pass,[14] but rather than the doping and Fermi level related explanation given by the authors the passing polarization was probably determined by the superstrate refractive index. Consistent simulation results were obtained with the waveguide geometries presented by Kim and Choi (and for a fixed Fermi level of 100 meV), in qualitative agreement with the experiment. Quantitative agreement would, however, require more information about the exact waveguide cross section and about the presence of multilayer graphene in the reported experiments.[14]

The conclusions drawn here also explain the experimental results reported by Lee *et al.* in the optical fiber geometry[15]. Lee *et al.* show a TM-pass polarizing behavior in a D-shaped optical fiber covered with graphene, obtaining an attenuation of 0.27 dB for the TE radiation and of 0.20 dB for the TM radiation, over a length of 5 mm. Compatible values were obtained in our simulations: 0.29 dB for the TE radiation and 0.16 dB for the TM radiation. Lee *et al.* also place a liquid with a refractive index of 1.423 over graphene and report an attenuation increase for the TE radiation to 3.47 dB, while the attenuation of the TM radiation reduces to 0.11 dB; our simulations under the same conditions resulted in attenuations of 3.35 dB, for the TE radiation, and 0.11 dB, for the TM radiation, in good agreement with the experimental results. The TM radiation transmission enhancement after the liquid superstrate addition is explained by Lee *et al.* as being the result of a reduction in the TM radiation scattering, due to the reduction in the refractive index contrast between fiber core and superstrate. However, the good agreement with our simulations, which do not account for scattering, indicates otherwise. Here, we explain the enhancement in TM radiation transmission as a result of the reduction in the fraction of the electric field that is parallel and overlaps with graphene (i.e., the reduction of $U_{\parallel}$).

Among the simulated waveguides the only significant discrepancy with experiments was obtained for the work reported by Bao et al.,[11] that explores an optical fiber geometry. It reports a TE-pass behavior, in contrast to the TM-pass behavior expected from our simulations and also observed by Lee et al.[15] in a similar geometry. The simulation, performed using the waveguide parameters provided by Bao et al. with 3 mm graphene length,[11] predicts absorptions of 0.24 dB for the TE radiation and 0.12 dB for the TM radiation, but the reported values were ~21 dB and ~45 dB respectively for the TE and TM radiation.[11] The high values of optical losses disagree with both our simulations and the results reported by Lee et al.[15] and, thus, extra losses (e.g., from scattering) might have interfered in the experimental results.[11] Indeed, it is possible that a rough fiber and/or graphene surface resulted in the higher scattering of the TM-radiation causing the observed polarizing effect.[27] Bao et al. report that the graphene film in their experiment is discontinuous with many gaps and wrinkles,[11] which may lead to scattered light. The propagation of the TE radiation with a lower loss is explained by Bao et al. as being a consequence of the propagation of SPPs occurring in the vicinity of the Fermi level corresponding to half the photon energy value, due to the negative imaginary part of graphene's conductivity,[28] see Fig. 2. However, the TE-pass behavior is reported for different wavelengths (488 nm, 532 nm, 980 nm, 1300 nm, 1480 nm, 1550 nm[11]) in the same device, without any tuning of the Fermi level. Therefore, the higher loss of the TM radiation is more likely to have occurred due to scattering or polarization-selective light coupling losses. We add that the model developed by Bao et al. to describe their results simplifies the waveguide geometry, converting the optical fiber into a slab waveguide. However, as shown here, the waveguide design crucially determines the tangential optical field that overlaps with the graphene layer, and such a geometry simplification can significantly affect the characteristics of the graphene-waveguide polarizer.

The results and analysis developed here for single-layer graphene can be directly extended for stacked few-layer graphene. Considering a weak interaction between graphene layers, the net optical conductivity can be considered, in this case, to be the sum of the contribution of each layer, which is found to lead to loss coefficients and extinction ratios (in dB) that are those of a single layer multiplied by the number of graphene layers. This is in agreement with the results reported by Lee et al.[15] for two stacked graphene layers, which doubled the extinction ratio compared to single-layer graphene. However, experimental results deviate from the linear behavior for four layers of graphene,[15] which may be attributed to a reduction in the quality of the stacked graphene sample, inducing increased scattering losses. Light scattering may also explain the small influence of the number of graphene layers on the results reported by Bao et al.,[11] with an extinction ratio increase by only 1 dB with 5 layers compared to the single layer.

Besides optical scattering caused by wrinkles, cracks and contamination on graphene, the polarizing effect can, in principle, be affected by scattering due to the waveguide roughness.[27] However none of the experimental results reported[11,13-15] seem to have a significant polarization dependent loss in the waveguide before the graphene transference. Therefore, the main challenge for the fabrication of graphene-waveguide polarizers appears to lie in the quality of the graphene and in the transference method for obtaining a smooth, low contamination and continuous graphene layer.

Graphene-based optical polarizers can be affected by graphene's nonlinear optical response at high intensities. At this regime saturable absorption[3,5] can lead to a smaller polarization-dependent loss, reducing the polarizing effect. Optical intensities of 7.89 MW cm$^{-2}$ have been reported to be required for saturation to be observed at 1550 nm[5]. However, considering that only a portion of the light interacts with graphene through its evanescent field, higher intensities can propagate along the waveguide before any observable saturation.

Finally, our analysis can be extended to different wavelengths; simulations performed at 532 nm and 1064 nm wavelengths resulted in the same dependence of the polarizing effect on the waveguide design, provided that the waveguides is properly scaled. This scale dependence can be exploited to create wavelength-dependent polarizers.

In conclusion, we showed that the waveguide design, rather than graphene's Fermi level, as widely believed, determines the polarization properties of graphene-based waveguide polarizers because it defines the fraction of the modal electric field that is parallel and superposed with graphene. More specifically, we showed that graphene-waveguide polarizers can work either as TE or TM pass devices depending on the waveguide dimensions and the superstrate refractive index. The development of optimized polarizers must, therefore, take into account a careful design of the waveguide, which is usually either overlooked or taken into account solely for increasing the light-graphene superposition. We stress that simply increasing this superposition does not suffice for maximizing the polarization extinction ratio, as the electric field perpendicular to graphene is not absorbed. The results presented here establish the foundations for the development of optimized graphene-waveguide polarizers and explain their operation.

## Methods

**Simulation.** The modal analyses were performed in the COMSOL Multiphysics. The number of degrees of freedom solved was 1,047,000 using adaptive triangular meshes with maximum element size of 80 nm and minimum element size of 0.12 nm in the region around the graphene.

**Waveguide Parameters.** The silicon waveguide reported by Kou et al.[13] was simulated with the following parameters: $n_1$=3; $n_2$=$n_{top}$=1; $n_{sub}$=1.45; $W$=400 nm; $H$=200 nm; and light wavelength 1550 nm. The convex polymeric waveguide reported by Kim and Choi[14] was simulated with the parameters: $n_1$=1.39; $n_2$=$n_{top}$=1 or 1.37, for the cases without or with the cladding respectively; $n_{sub}$=1.37; width at half the maximum height was assumed $W$=5.75 μm with 68° wedge angle; height $H$=4.8 μm and 1310 nm wavelength. The D-shaped optical fiber based polarizer reported by Bao et al.[11] was simulated with core and cladding refractive indices of 1.449 and 1.444 respectively; core diameter 8.2 μm and wavelength 1550 nm. The flat section of the fiber was assumed to be polished down to 1 μm into the fiber core. The same optical fiber parameters were used for simulating the device reported by Lee et al.,[15] except for the distance from the polished region to the core edge, which was set to zero in this case; the superstrate region was simulated with a refractive index of 1 for air and of 1.423[15] for the liquid superstrate. Graphene was simulated as a boundary condition for the cases described here.

## Acknowledgements

This work was supported by FAPESP (grant. n° 2012/50259-8), CNPq, and MackPesquisa.

## Author contributions statement

R. E. P. de O. performed the simulations. R. E P. de O. and C. J. S. de M. performed the analysis and elaborated the manuscript.

## Additional information

**Competing financial interests:** The authors declare no competing financial interests.